**Estimating direction in brain-behavior interactions: Proactive and reactive brain states in driving**

Javier O. Garcia[1], Justin Brooks[1], Scott Kerick[1], Tony Johnson[5], Tim Mullen[2], & Jean M. Vettel[1,3,4]

[1]U.S. Army Research Laboratory, Aberdeen Proving Ground, MD
[2]Qusp Labs., San Diego, CA
[3]University of California, Santa Barbara, CA
[4]University of Pennsylvania, Philadelphia, PA
[5]DCS Corporation, Alexandria, VA

Highlights
• Traditional neuroscience studies investigate localized task-evoked responses
• Our approach examines continuous tracking of brain-behavior interactions in oscillatory activity
• Brain leads behavior in a Proactive state, while brain follows behavior in a Reactive state
• Reactive states are largely carried by alpha activity in regions sensitive to environmental statistics
• Proactive states rely more on a diffuse delta-beta network, particularly when linked with steering behavior

Keywords: EEG, driving, neuro-behavioral analysis, source analysis,

Running Head: Neuro-behavioral state modulations in driving

*Corresponding Author:
Javier O. Garcia
Email: javier.o.garcia.civ@mail.mil
RDRL-HRS-C (ARL/HRED/TNB)
459 Mulberry Point Road
Aberdeen Proving Ground, MD 21005

**Abstract**

Conventional neuroimaging analyses have revealed the computational specificity of localized brain regions, exploiting the power of the subtraction technique in fMRI and event-related potential analyses in EEG. Moving beyond this convention, many researchers have begun exploring network-based neurodynamics and coordination between brain regions as a function of behavioral parameters or environmental statistics; however, most approaches average evoked activity across the experimental session to study task-dependent networks. Here, we examined on-going oscillatory activity and use a methodology to estimate directionality in brain-behavior interactions. After source reconstruction, activity within specific frequency bands in *a priori* regions of interest was linked to continuous behavioral measurements, and we used a predictive filtering scheme to estimate the asymmetry between brain-to-behavior and behavior-to-brain prediction. We applied this approach to a simulated driving task and examine directed relationships between brain activity and continuous driving behavior (steering or heading error). Our results indicated that two neuro-behavioral states emerge in this naturalistic environment: a *Proactive* brain state that actively plans the response to the sensory information, and a *Reactive* brain state that processes incoming information and reacts to environmental statistics.

**Graphical Abstract**

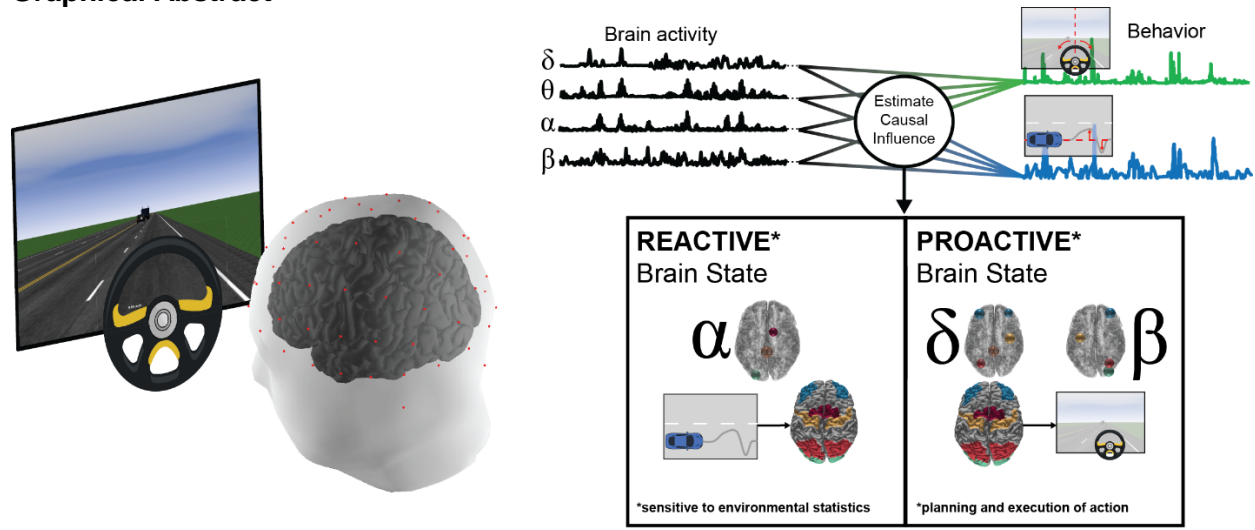

**Introduction**

The brain is composed of roughly 160 billion neural and non-neural support cells that coalesce into dynamic, neuronal assemblies of coordinated activity (Azevedo et al., 2009), and neuroscientists have developed a battery of neuroimaging and analysis techniques to study the local specialization of neuronal populations at the macro-scale level of organization. In EEG, brain activity has often been explored with event-related potential (ERP) analyses that reveal localized peak responses in scalp electrodes that differentiate experimental conditions (for review, see Luck, 2014) or by averaging EEG data organized into epochs and reconstructed in source space to examine condition-specific effects in more localized regions of interest in a 3D head model (Michel et al., 2004). Likewise, conventional fMRI analyses have subtracted whole-brain activation between experimental conditions to identify localized regions of task-specific computational processing or analyzed *a priori* regions of interest, defined functionally or anatomically, to quantify their sensitivity to a variety of stimuli or task demands (for review, see Huettel et al., 2004). Collectively, these imaging approaches have produced a rich understanding about segregated areas of the brain and local specialization within brain regions. Recently, however, there has been increased interest in examining how the brain coordinates activity across these spatially disperse regions (Alivisatos et al., 2012).

Researchers have developed several new methods to investigate entire brain networks, including diverse approaches such as independent component analysis (e.g., Calhoun et al., 2009), encoding and decoding algorithms (e.g., Serences and Saproo, 2012) and graph theoretical approaches (e.g., Bassett and Bullmore, 2006; Bullmore and Sporns, 2009). These advances have allowed us to investigate the symphony of neural processing rather than a compartmentalized snapshot of the brain's dynamic response, and research on brain connectivity among regions continues to increase within the field (e.g., Li et al., 2009; Sakkalis, 2011). Overall, connectivity-based neuroimaging methodologies show promise for augmenting our understanding of how dynamic changes in brain networks support millisecond fluctuations in behavior (Alivisatos et al., 2012; Friston, 1994; Sporns, Chialvo, Kaiser, & Hilgetag, 2004).

In particular, our research effort presupposes that functional network connectivity across disparate brain regions yields brain activity that underlies cognition and interaction in a complex world. We are interested in quantifying ongoing brain network dynamics that underlie the concept of a brain state, or the "fundamental algorithm by which cognition arises" (Gilbert and Sigman, 2007). Often termed state dependency, some researchers have investigated how resting state brain activity modifies incoming information (Wörgötter et al., 1998) or disrupts behavioral performance (Silvanto et al., 2008). Synchronized frequency oscillations are posited as a mechanism to form transient networks that can integrate information across local, specialized brain regions (He et al., 2015; Klimesch et al., 2007), and global brain dynamics characterized by specific frequency oscillations appear to have functional consequences (Buzsáki and Draguhn, 2004). Researchers have identified brain network responses related to task demands (DeSalvo et al., 2014)**,** stimulus properties (Stansbury et al., 2013)**,** and biomarker development for disease (Bassett et al., 2008)**.** In this study, we are interested in the relationship between distributed brain activity and continuous task performance, and we use EEG to study whole-brain oscillatory activity and capitalize on its temporal precision (Buzsáki and Draguhn, 2004; Engel et al., 2001; Steriade, 2001).

Here, we investigate temporal dynamics of brain activity and continuous behavioral performance in a simulated driving task. Driving is a complex visuo-motor task that requires interaction among cognitive systems to successfully navigate from one location to another, keep a safe distance from other vehicles, and maintain a consistent lane location while in motion. Despite the complexity of the task, experienced drivers successfully perform this task, with ease, often in a near-automatic fashion. Studies have investigated the underlying neural mechanisms in both real (Sandberg et al., 2011) and simulated (Calhoun et al., 2002; Spiers and Maguire, 2007) driving environments, explored networks that produce task failures (Simon et al., 2011), and used neural measures to predict vehicle parameters (Lin et al., 2005). Thus, a simulated driving task affords the opportunity to study relationships between continuous behavioral measurements and brain dynamics in a naturalistic, everyday task.

Our neuro-behavioral analysis method calculates time-varying asymmetries between fluctuations in oscillatory activity and two measures of driving behavior. Oscillatory activity is estimated for four common frequency bands within 12 cortical regions of interest (ROI) determined *a priori* from previous driving research (Calhoun et al., 2002; Spiers and Maguire, 2007). The two behavioral measures, steering wheel angle and vehicle heading error, were chosen because previous research has suggested that heading error of the vehicle is used to determine the steering response and tightly coupled with brain dynamics (Hildreth et al., 2000; Li and Cheng, 2011). In this framework, the heading error is a kinematic variable used to scale a steering response, and the steering response relates back to the heading error by a dynamic transfer function that accounts for vehicle speed and current heading, among other parameters. In this manner, examining the neural dynamics related to heading error and steering wheel angle permit the investigation of the brain's "closed loop" control of the vehicle. From this analysis, we identify two distinct neuro-behavioral brain states: a *Proactive* state where the brain activity predominantly causes behavior and a *Reactive* state where the brain activity is predominantly caused by behavior. We apply analysis of variance to investigate the effects of ROI, frequency band, and experimental factors on the proportion of time spent in each state, as well as the transition probability within and between states. Our analysis suggests that the Proactive state actively plans the response to the sensory information and the Reactive state processes incoming information and reacts to statistics of the environment.

**Method**

    Twenty-eight neurologically healthy volunteers participated in this experiment. This study was conducted in accordance with IRB requirements (32 CFR 219 and DoDI 3216.02). Upon arrival to the lab, participants were introduced to the driving environment and instructed how to perform the task. Subjects were asked to maintain the vehicle in the center of the rightmost lane of a four-lane highway (two lanes in each direction) and to maintain consistent vehicle speed at 45 mph as precisely as possible (See Figure 1 for a diagram of the display). Lateral perturbations resembling wind gusts were periodically imposed on the vehicle causing changes in its heading, and the participants were instructed to counter them by steering the vehicle back into the center of the rightmost lane as quickly and accurately as possible. Training on the task consisted of participants driving for 10-15 min until asymptotic performance in steering and speed control was demonstrated. They were then outfitted and prepped for the EEG acquisition. Following completion of the training and experimental setup, the participants proceeded to drive in a 45-min experimental condition where traffic density was manipulated (sparse, heavy). Vehicle perturbations ('wind gusts') were also presented in blocks of either high (every 8-10 s) or low (every 24-30 s) rates. These manipulations were introduced to make the driving experience more naturalistic and to investigate whether either factor imposed a modulation on the measured neuro-behavioral states.

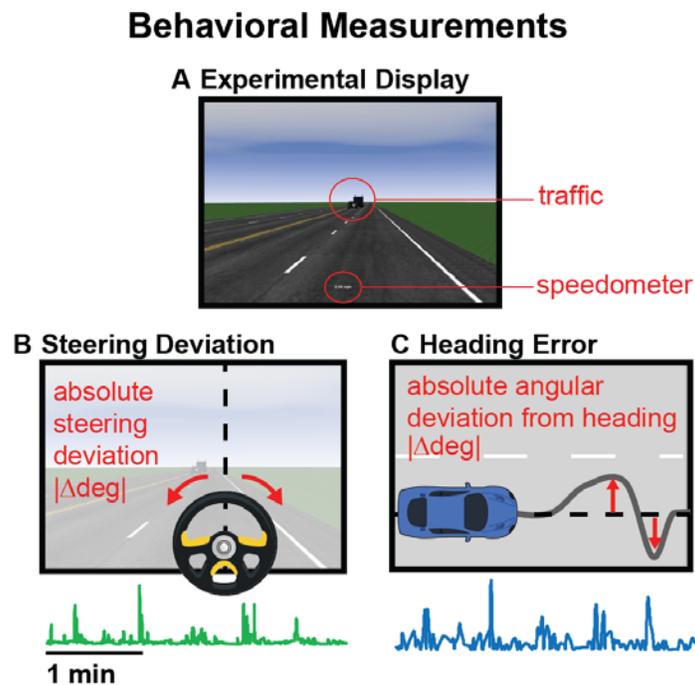

**Figure 1:** Experimental display and behavioral measurements. A) In this experiment, subjects were asked to maintain course in the far-right lane whilst on-going traffic and perturbations were introduced. A speedometer reading in the center bottom of the display indicated current speed. B) Continuous steering deviation, the absolute angular difference from the stationary angle (deg). Lower time course is the steering deviation from approximately 3 min of the experiment for one participant. C) Heading error, the absolute angular deviation the vehicle was positioned from the center of the right lane (deg). Lower time course is the heading error from approximately 3 min of the experiment from one participant.

*Neuro-behavioral Analysis*

An overview of the analysis steps are graphically described in Figures 2 and 3 and succinctly introduced here. First, standard preprocessing of EEG was completed on the raw signal, and continuous behavioral measures were temporally resampled and synchronized with the EEG signal. Next, cortical current source density (CSD) was estimated using cortically constrained low resolution electrical tomographic analysis (cLORETA), and the mean CSD was obtained for 12 regions of interest (ROI) defined *a priori* from previous literature. For each ROI, the Hilbert transform was applied to obtain spectral analytic amplitude within four common frequency bands (delta, theta, alpha, and beta). For each band, time-varying dependency between spectral amplitude and each behavioral measurement (steering and heading error) was inferred using generalized partial directed coherence (GPDC). To obtain the GPDC, we fit non-stationary multivariate autoregressive (MVAR) models to the data using dual extended Kalman Filtering. From GPDC estimates, a measure of asymmetry was obtained and used to assign *Proactive* vs. *Reactive* state labels to each time point. Finally, we statistically analyzed the effects of cortical ROI, frequency band, and experimental traffic manipulations (traffic density and vehicle perturbation frequency) on the proportion of time spent in each state as well as on the transition probability within and between states. Each step of this pipeline is described in further detail below.

## Data Acquisition and Preprocessing

EEG measurements were made using a 64-channel (1024Hz sampling rate) Biosemi ActiveTwo System (Biosemi Instrumentations, The Netherlands). Raw EEG measurements were pre-processed using in-house software in Matlab (Mathworks, Inc.) and the EEGLAB toolbox (Delorme and Makeig, 2004). The pre-processing pipeline largely follows the PREP approach (Bigdely-Shamlo et al., 2015) and contains five steps: (1) resampling the raw EEG to 250Hz, (2) line noise removal via a frequency-domain (multi-taper) regression technique to remove 60Hz and harmonics present in the signal, (3) a robust average reference with a Huber mean, (4) artifact subspace reconstruction to remove residual artifact (the standard deviation cutoff parameter was set to 8), and (5) a piece-wise detrending algorithm to remove low frequency drift in the signal (window size = 312ms, step size= 20ms).

We computed two measures of driving behavior: steering deviation as the absolute angular rotation of the steering wheel (measured in degrees), and heading error as the absolute angular deviation of the vehicle's motion trajectory and a line parallel with the simulated, indefinitely straight road (measured in degrees). These synchronized behavioral measures (100Hz sampling rate) were recorded using a distributed architecture, in which multiple data streams were recorded by different CPU's via an Arduino-based system (Brooks and Kerick, 2015; Jaswa et al., 2012). Each computer in the system produced data logs that included the common sync marker, and synchronization was performed in post-processing. The measured jitter within the system was confirmed to be below the resolution of the analysis (50Hz). These behavioral measures were then converted to degrees, and the absolute value was taken for the neuro-behavioral analysis.

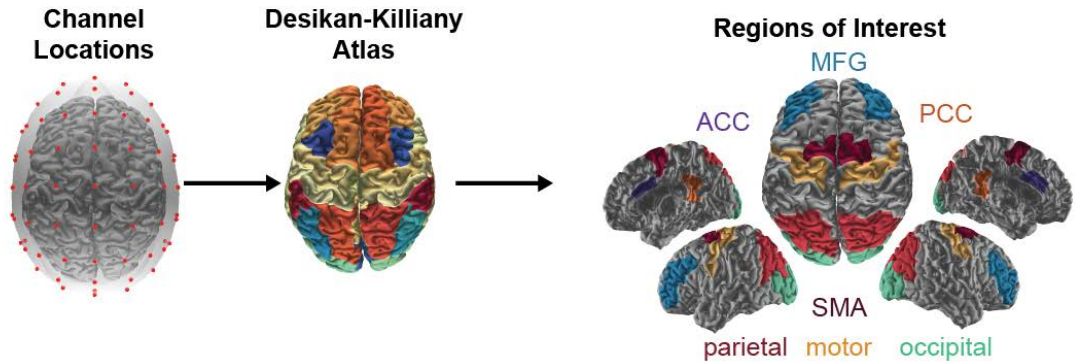

**Figure 2**: Regions of interest, channel locations, and atlas demarcation. Standard 64 channel EEG channel positions and a headmodel created from the Colin brain in MNI space was used to transform the EEG channel data into cortical current source density (CSD) via cLORETA. For subsequent analysis, CSD was averaged within 12 ROIs: the posterior and anterior cingulate cortices (ACC/PCC) as well as bilateral regions in middle frontal gyrus (MFG), supplemental motor area (SMA), parietal cortex, motor cortex, and lateral occipital cortex.

Distributed Source Reconstruction

From the pre-processed EEG data, we estimated current source density over a 5003-vertex cortical mesh. A boundary element method (BEM) forward model was derived from the 'Colin 27' anatomy (Holmes et al., 1998) and transformed into MNI305 space (Evans et al., 1993) using standard electrode positions fit to the Colin 27 head surface in BrainStorm (Tadel et al., 2011). The BEM solution was computed using OpenMEEG (Gramfort et al., 2010; Kybic et al., 2005), and the cLORETA approach was used for inverse modeling as described in detail in (Mullen et al., 2015) and implemented in the BCILAB (Kothe and Makeig, 2013) and Source Information Flow (SIFT)(Mullen, 2014) toolboxes.

Using averaged CSD from appropriate vertices of the cortical mesh, the functional activity in 12 *a priori* ROIs selected from previous driving studies (Calhoun et al., 2002; Spiers and Maguire, 2007) were estimated from the 5cm (496 parcel) subparcellation of the Desikan-Killiany atlas (Desikan et al., 2006). These ROIs were anterior cingulate cortex (ACC), posterior cingulate cortex (PCC), and bilateral regions corresponding to middle frontal gyrus (MFG), supplementary motor area (SMA), parietal cortex (portions of the inferior and superior parietal lobe), motor cortex (dorsal precentral gyrus), and lateral occipital regions. ROI locations on the mesh are visualized and labeled in Figure 2.

Power Spectral Estimation

Continuous, time-varying measures of spectral power within delta (2-3Hz), theta (4-7Hz), alpha (8-12Hz), and beta (13-25Hz) frequency bands were obtained for each ROI. For each frequency band, the time-series were filtered with a zero-phase FIR band-pass filter with 6dB attenuation, as implemented in EEGLAB (Delorme and Makeig, 2004), then Hilbert transformed to extract the complex analytic amplitude, and finally the magnitude-squared instantaneous power was obtained. Since power and behavioral measures change relatively slowly, to improve computational efficiency of subsequent modeling steps and reduce model complexity, these measures were downsampled to 50Hz prior to modeling.

Neuro-behavioral Relationships

As graphically outlined in Figure 3, time-varying dependencies between ROI power and behavioral measures were inferred using an effective connectivity measure related to Granger-Geweke causality (Geweke, 1982; Granger, 1969). When time series data are fit by a multivariate autoregressive (MVAR) model by minimizing the error terms, the model coefficients provide information about time lag influences between the signals and capture the causal dependencies between two or more time-series. However, brain and behavioral dynamics are typically non-stationary (Boashash et al., 2000; Ku and Kawasumi, 2007).To account for this, we modeled non-stationary brain-behavior dynamics with a locally-linear MVAR dynamical model estimated using dual extended Kalman filtering (DEKF) (Wan and Nelson, 1997).This method has previously been used to model non-stationary causal influences in EEG data (Omidvarnia et al., 2011).

Using the SIFT toolbox, a 5th order time-varying MVAR model was fit to each pair of normalized bandpower and behavioral measure time-series using DEKF. The DEKF forgetting factor was set to .01, allowing an effective time window of 2 seconds. The MVAR model coefficients were then used to estimate generalized partial directed coherence (GPDC, Baccalá and de Medicina, 2007). This was integrated over all frequencies to yield a time-domain GPDC causal estimate ($GPDC_{net}$).

The GPDC has the important property of being scale-invariant, ensuring that the inferred strength of the granger causal influence is independent of the relative scale of data time-series. Furthermore, since MVAR modeling assumes homoscedastic (equal variance) data, power and behavioral time-series were temporally normalized using an adaptive z-scoring method within 2 sec windows prior to model fitting. Levene's test for homoscedasticity was applied within a 2 sec sliding window (step = 1.5 sec) to confirm equal variance between modeled time-series. Results showed that on average, across subjects, frequency bands, and behavioral measures, only 1.8% of windows showed significant differences in variance between power and behavior time-series pairs (p < .05, FDR corrected) confirming the effectiveness of the adaptive z-scoring procedure.

## Sample subject analysis pipeline

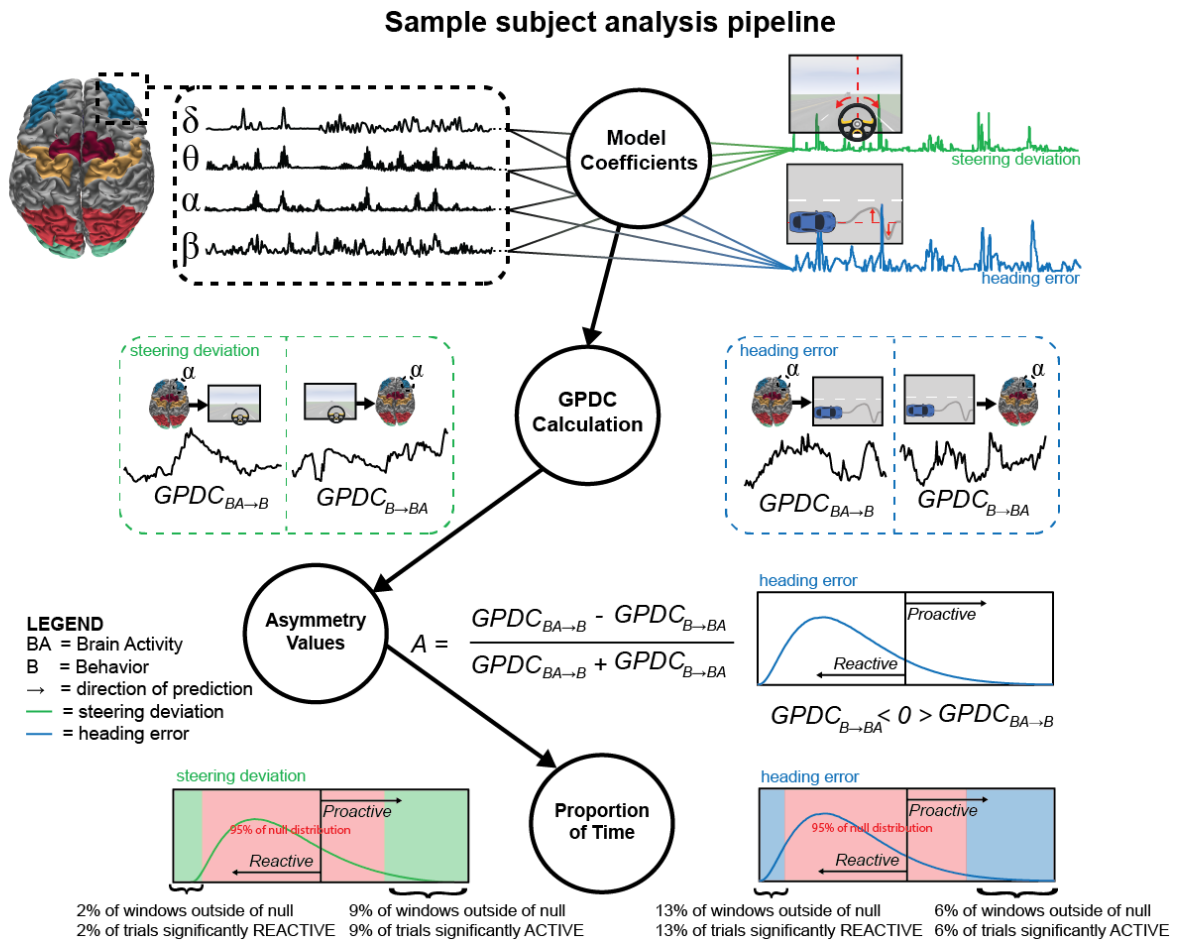

**Figure 3:** Graphical display of analysis. Power in each frequency band was obtained for each brain region. Pairwise MVAR models were fit to each pair of EEG frequency band and behavioral measure (Model Coefficients) using a dual extended Kalman filter (DEKF) and Generalized PDC was calculated (GPDC Calculation). Asymmetry was defined as the difference divided by the sum of the GPDC measures from brain activity to behavior and vice versa (Asymmetry Values). The final step was to compare neuro-behavioral states that significantly deviated from a null distribution obtained by shuffling the timecourses for behavior and brain activity (Proportion of Time). This allowed us to create a measure that corresponded to the amount of time each participant's brain was significantly in a Proactive or Reactive brain state.

GPDC estimates of casual influence were calculated for each time-series pair (12 regions, 4 bands, 2 behaviors = 96 $GPDC_{net}$ time courses per subject). We then summarized the granger causal influence of brain activity (BA) on behavior (B) ($GPDC_{BA \to B}$) or relative to that of behavior on brain activity ($GPDC_{B \to BA}$) using the asymmetry ratio:

$$A = \frac{GPDC_{BA \to B} - GPDC_{B \to BA}}{GPDC_{BA \to B} + GPDC_{B \to BA}}$$

Thus, when $A$ is greater than zero, the relative influence of brain activity on behavior is greater than the influence of behavior on brain activity (Figure 3, Asymmetry row). We label these states as ***Proactive*** since the neural activity can be thought of as controlling the driving behavior. Conversely, when $A$ is less than zero, behavior predominantly influences brain activity, and we label these states ***Reactive*** since the neural activity can be thought of as occurring in response to actions taken in the driving task.

Next, for each frequency band, ROI, brain state, and behavioral measure, we calculated the proportion of asymmetry values larger in magnitude than expected under a null hypothesis of granger causal independence of brain and behavioral measures. For each pair of time-series, an asymmetry null distribution, $A_{null}$, was constructed. This destroys the temporal dependency between time-series by circularly shifting the continuous time-series of each behavioral measure relative to each corresponding brain measure. This was followed by GPDC and asymmetry ratio calculation as described above. Asymmetry values lying outside the central 95% of $A_{null}$ were then deemed significant at the level of $p$<0.05. Example null distributions are depicted in Figure 3, bottom row.

Analyses of variance (ANOVAs) were applied to proportional values to quantify the effect of experimental and neurophysiological factors on the proportion of time spent in Proactive and Reactive states for steering and heading error. The factors examined were (1-2) two experimental traffic manipulations, traffic density and perturbation frequency,(3) neuronal frequency band and (4) cortical ROI.

Finally, we investigated the relative stationarity of Proactive and Reactive states for each behavioral measure and frequency band by estimating the transition probability between and within each state: Proactive-to-Proactive, Proactive-to-Reactive, Reactive-to-Reactive, Reactive-to-Proactive. The transition probability $p_{ij} = P(S_t = j \,|\, S_{t-1} = i)$ , where $S_t$ is the state at time $t$ and $i,j \in \{$Proactive, Reactive$\}$,was obtained using the maximum likelihood estimate $p_{ij} = n_{ij}/\sum_j n_{ij}$, where $n_{ij}$ is the number of sequential transitions from state $i$ to $j$. We then submitted these transition probabilities to an ANOVA with 12 regions, 4 frequency bands, 2 behaviors, and 4 state transitions as factors.

Post-hoc paired $t$-tests were used to investigate simple effects driving main effects and interactions found for each ANOVA. All $t$-tests are reported with significance corrected for multiple comparisons with the false discovery rate procedure (FDR, Benjamini and Yekutieli, 2001).

## Results

In this experiment, we study the relationship between the temporal dynamics of EEG oscillatory activity and continuous fluctuations in two behavioral measures of driving performance in a simulated environment, the participant's steering behavior and vehicle's heading error (Hildreth et al., 2000; Li and Cheng, 2011). Classical granger causal estimates of connectivity assume a stationary and linear representation of multichannel or multisource EEG activity, an assumption that is often inaccurate when modeling EEG data. Here, we estimate the cortical source activity, parcellate mesh vertices using the Desikan-Killiany atlas (Desikan et al., 2006), and select 12 *a priori* regions of interest from previous driving (Calhoun et al., 2002; Spiers and Maguire, 2007). We then use a time-varying model, a dual extended Kalman filter (DEKF), to link brain activity to behavior, and we calculate generalized partial directed coherence, GPDC, to estimate the granger causal influence for each ROI-behavior pair in four frequency bands (delta, theta, alpha, and beta). Thus, this analysis computes a total of 96 GPDC time courses per participant from 12 *a priori* regions, 4 frequency bands, and 2 continuous driving behaviors.

Our neuro-behavioral analysis uses an asymmetry measure to emphasize the directionality of the casual GPDC brain-behavior relationship during the driving task. When values are higher for $GPDC_{BA \rightarrow B}$, the brain activity precedes and predicts behavior performance, and we label these time intervals as a Proactive brain state since the neural activity can be thought of as controlling the driving behavior. Conversely, when values are higher for $GPDC_{B \rightarrow BA}$, the behavioral performance precedes and predicts brain activity, and we label these time intervals as a Reactive brain state since the neural activity occurs in response to actions needed in the driving task. The significant time intervals of the brain state in these time intervals is computed by comparing to a permuted null distribution of the brain-behavior values, $GPDC_{null}$. This asymmetry measure is used for two of the result sections: the first identifies the dominant brain state by quantifying the distribution of asymmetry values, and the other estimates the transition probability of switching between brain states.

For the other three result sections, we calculate the proportion of time each of 96 GPDC time courses was in the Proactive or Reactive brain states to examine differential relationships among regions, frequency bands, and the two driving performance measures. These proportion values were used to examine (1) the effect of two experimental traffic manipulations, traffic density and perturbation frequency, on detected brain states, (2) the dependence of brain states on particular frequency band oscillations, and (3) the regions that make up the brain networks that underlie the Proactive and Reactive brain states in this simulated driving task.

*Subjects are primarily in a reactive state*

First, we examine histograms of the asymmetry measure to assess the relative frequency of Proactive and Reactive states in the simulated driving task (Figure 4). We observe that group-averaged histograms for both steering and heading error are left-skewed. This suggests subjects were primarily in a Reactive brain state with behavior predominantly influencing brain activity, for each subject (mean = .84 vs .16, SD = .03, across subjects). Further, for a subset of the subjects, there was an effect between driving measures within the Proactive brain state. Figure 4 shows a larger proportion of time spent in the Proactive state (right side of histogram, positive asymmetry values) for steering (green) versus heading error (blue).This likely reflects demand characteristics of the driving measurement since steering is arguably a more *proactive* behavioral estimate than heading error.

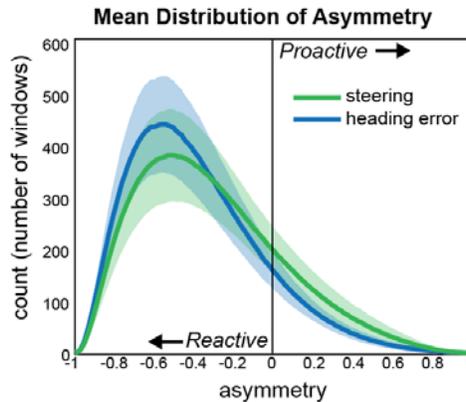

**Figure 4:** Histograms of asymmetry values across the experiment, averaged over all subjects (light shaded regions denote 1 standard deviation across subjects).

*Effects of task manipulations on neuro-behavioral states*

Although the Reactive state is most common in this simulated driving task, we ask whether the proportion of time spent in Reactive and Proactive states reflect task demands in driving. The experimental design incorporated two major task manipulations during the 45-min highway drive: traffic density was blocked as intervals of sparse and heavy traffic, and the frequency of vehicle perturbations ('wind gusts') occurred in blocks of either high (every 8-10 s) or low (every 24-30 s) rates. These manipulations were introduced to make the driving experience more naturalistic and to investigate whether either factor imposed a modulation on brain and behavioral measures. For example, more frequent perturbation events during driving may require a more frequently Proactive state to ensure efficient correction to lane deviations.

To examine the effect of the traffic manipulations, we submitted the proportion of time participants are within each neuro-behavioral state to an ANOVA with brain state, traffic density, and perturbation frequency as factors. First, only a main effect of brain state was found ($F(1,222)$ = 17.6, $p < .001$), confirming the analysis of histogram values (Figure 4) that identified a predominance of the Reactive brain state in this simulated driving task.

Second, two significant interaction effects were found: the first was a state by traffic density interaction ($F(1,222)$ = 4.5, $p = .03$), and the second was a brain state by perturbation frequency interaction ($F(1,222)$ = 4.5, $p = .04$). As Figure 5 shows, the state by traffic density interaction was driven by a reversal of effects for the traffic manipulation: in the high traffic density condition, there is more time spent in the Reactive state and less time spent in the Proactive state, whereas the amount of time is similar across the two brain states in the low traffic density condition. The brain state by perturbation frequency was driven by an increase in amount of time in a Reactive state between low and high perturbation frequency conditions, whereas the amount of time in a Proactive state is slightly decreased between the low and high perturbation frequency. Collectively, these interaction effects suggest that the Proactive brain state is more sensitive to these experimental conditions than the Reactive state.

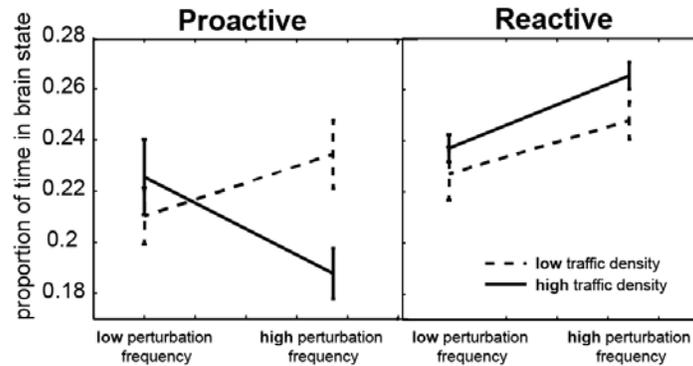

**Figure 5:** Effect of perturbation frequency on neuro-behavioral state. Lines represent the mean proportional time for the Proactive (left panel) and Reactive (right panel) brain states within two experimental conditions: perturbation frequency (columns, x axis) and traffic density (lines). Error bars are SEM.

*Switching between states*

We have shown that subjects are primarily in a Reactive state within this experiment and the Proactive state is more malleable to task demands. But how often do the participants switch to each brain state and what are the temporal dynamics of this transition between brain states? The asymmetry values from the right parietal cortex are shown in Figure 6A for approximately 3 minutes for a single subject's steering behavior with each of the 4 frequency bands. This visualization highlights our general observation that state transitions are infrequent but statistically significant timeframes occur for each neuro-behavioral state.

We quantify the transition probability between brain states using statistically significant asymmetry values. On average, the transition probability *within* states is .48 (Proactive-to-Proactive and Reactive-to-Reactive), whereas *between* states (Proactive-to-Reactive and Reactive-to-Proactive) is .02. These results indicate that transitions between Proactive and Reactive states are infrequent (Figure 6B). An ANOVA with driving behavior, state transition (Proactive-to-Reactive, Reactive-to-Proactive, Reactive-to-Reactive, Proactive-to-Proactive), and frequency band as factors showed main effects of state transition ($F_{(3,10364)} = 6.9 \times 10^5$, $p < .001$), reflecting the large difference between within state stationarity (.48) compared to the rare state transition (.02). The ANOVA also revealed two significant interactions, one between behavior and state transition ($F_{(3,10364)} = 39.9$, $p < .001$) and another between frequency band and state transition ($F_{(9,10358)} = 220$, $p < .001$).

As shown in Figure 6C, the behavior by state transition interaction was driven by the increased *within* state probability from Reactive-to-Reactive for heading error compared to steering deviation and an increase in transition probability for the steering behavior when inspecting *between*-state transitions (Reactive-to-Proactive, Proactive-to-Reactive). This suggests that the brain stays in a reactive state in conjunction with the heading error measure more frequently than with the steering angle. This result confirms the histogram analysis (Figure 4) that suggested the demand characteristics of steering may require the Proactive state compared to heading error.

To further examine the *between* state transitions (R-P and P-R) observed only in conjunction with the steering measure, we investigated frequency band by state transition interaction, and Figure 6D displays the means for the *between* state transitions. In FDR-corrected paired *t*-tests (q < .05) between each frequency pair, we find that all pair wise comparisons are significantly different from each other with alpha showing the highest probability of Proactive-to-Reactive and Reactive-to-Proactive state transitions.

Collectively, these results confirm the dominance of the Reactive brain state, particularly for heading behavior, and the largest transitions between the two brain states occur in the alpha band in conjunction with steering behavior which is an arguably more Proactive behavioral measure.

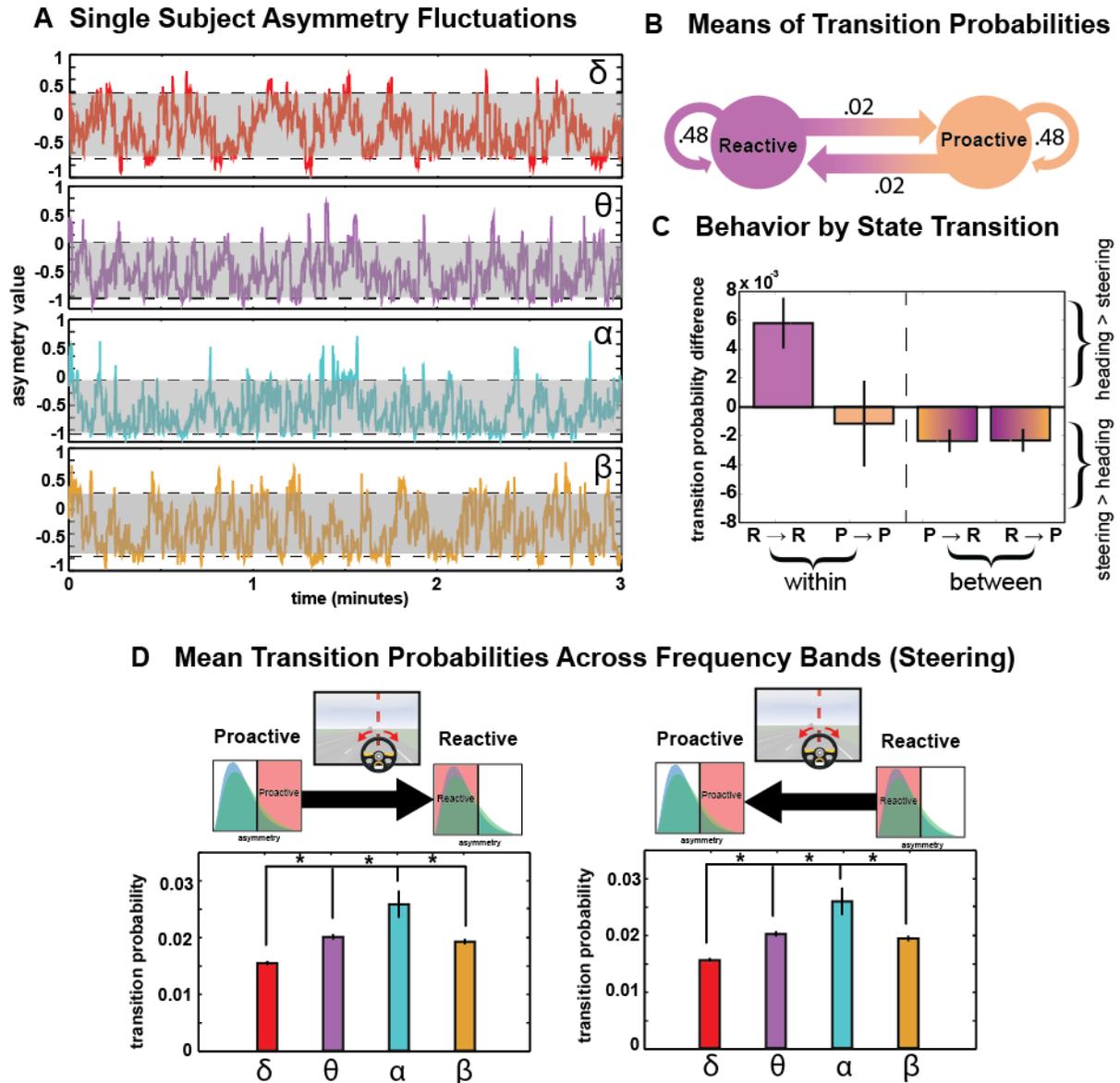

**Figure 6**: A) A 3-minute segment of continuous asymmetry values between parietal ROI power and steering for delta, theta, alpha, and beta bands. Significant asymmetry values lie outside the shaded regions, which denote the central 95th percentiles of corresponding null distributions. B) Mean transition probabilities reveal that transitions between Proactive and Reactive states are relatively infrequent. C) Mean difference (SEM across subjects) in neuro-behavioral state transition probability between the two behavioral measures (heading error – steering deviation) is shown. Positive values indicate greater transition probability for heading error, while negative values indicate greater transition probability for steering. For within state transitions, heading error has the most Reactive-Reactive transitions, while steering shows increased rate of transition between states for both Proactive-Reactive and Reactive-Proactive. D) The between state transition probabilities for steering behavior, averaged over ROIs, are shown for each frequency band. The probability of a state transition is significantly greater for Alpha than any other frequency band (q <.05, FDR corrected).

*Frequency interactions with neuro-behavioral state*

We continue to examine the proportion of time spent in Proactive and Reactive states as a function of frequency band. To quantify frequency effects, the proportions were submitted to an ANOVA with 12 regions, 4 frequency bands, 2 behaviors, and 2 brain states as factors. Main effects of frequency band (F(3, 5180) = 70.2, p < 0.001), behavior (F(1,5182) = 82.5, p < 0.001), and brain state (F(1,5182) = 660.2, p < 0.001) were found, with several significant interactions. Of note, behavior by brain state (F(1, 5182) = 155.0, p < 0.001) and frequency band by brain state (F(3, 5180) = 336.7, p < 0.001) were both significant.

To understand these effects, we performed post-hoc *t*-tests to examine pairwise differences in the mean proportion of time spent in each state, averaged over all ROIs and subjects, for all frequency bands, states, and behavioral measures (Figure 7). Significance tests were corrected for multiple comparisons using FDR (q < .05). The analysis revealed a number of significant differences in means. For the Proactive brain state, each of the four frequency bands showed statistically significant within-band differences in means between heading error and steering, further arguing that steering deviation is a more proactive measure. Furthermore, within each behavioral measure, the means were significantly different for all pairs of frequency bands. Conversely, for the Reactive state no significant differences in means were found within or between behavioral measures, although the alpha and delta comparison for steering is marginally significant (p < .05, uncorrected). The means for each frequency in the Reactive brain state in Figure 7 suggest that both heading error and steering deviation have similar frequency band profiles with a dominant role in the alpha band.

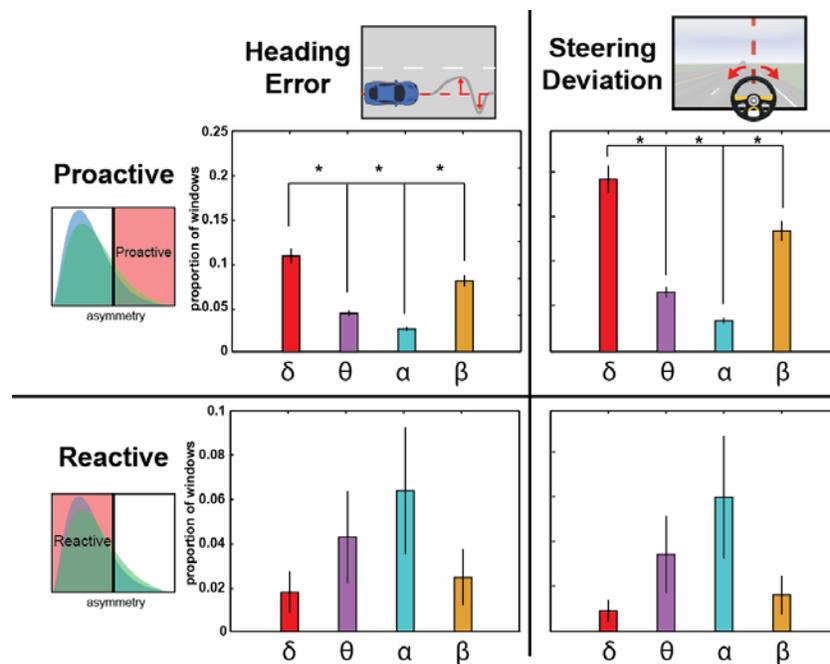

**Figure 7:** Proportion of time spent in Proactive and Reactive states, averaged over all ROIs and subjects, for each frequency band (delta, theta, alpha, beta) and behavioral measure (heading error, steering deviation). Error bars denote SEM. (*) denotes statistically significant differences in means at the FDR-corrected level of q < 0.05. For the Proactive state, the means are significantly different between all pairs of frequency bands. For the Reactive state, there were no significant differences in means between frequency bands.

*Regional contributions to neuro-behavioral state*

Finally, we investigate the regional contributions to Proactive and Reactive states as a function of frequency band. In Figure 8A, we first re-plot the frequency effects for steering that were shown in Figure 7 by placing a colored orb for each of our 12 ROIs (Figure 2), and its size represents the proportion of time measurement.

In Figure 8B, results from our final analysis are shown. Here, the orbs are scaled within those three frequency bands (mean subtracted) to reveal what regions are the strongest contributors for steering behavior across the two task states. For the Proactive state, steering behavior is dependent on delta activity across a diffuse set of brain regions consisting of SMA, motor, frontal, and PCC. For beta activity, steering behavior depends on right lateralized parietal, occipital, and frontal regions with substantial contribution from the motor cortex. For the Reactive state, steering behavior predominantly influences alpha activity in right parietal, motor, and frontal brain regions.

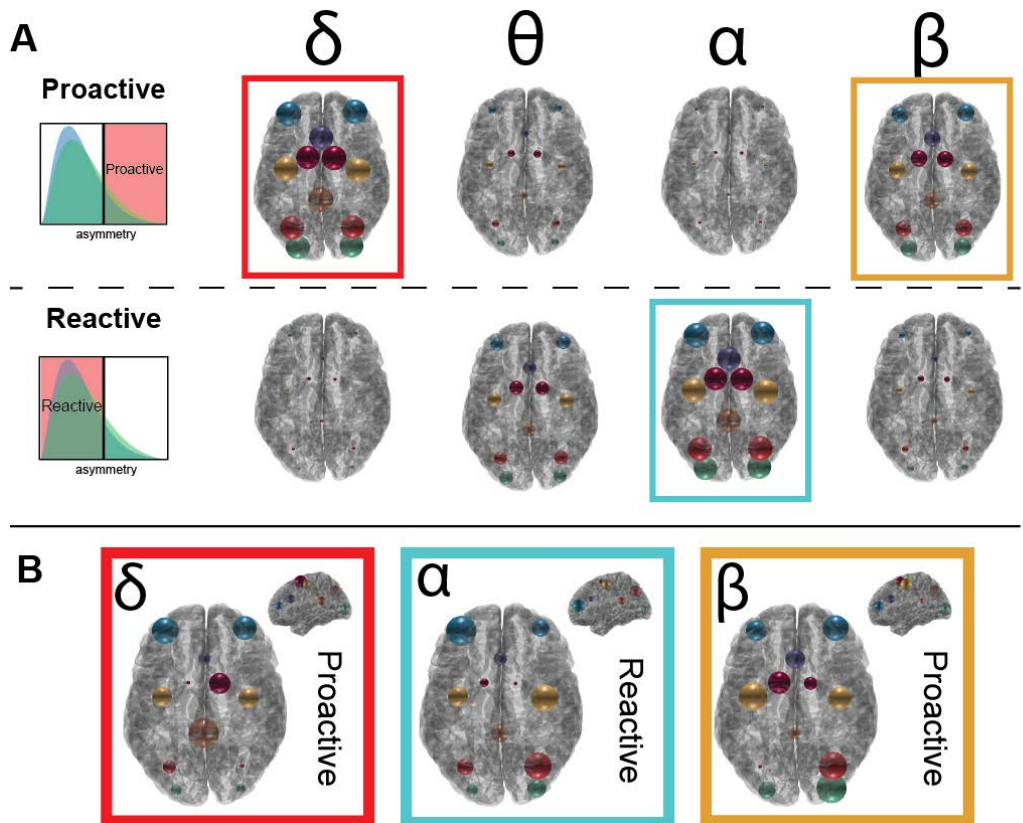

**Figure 8:** Data from Figure 7 (right panel; steering deviation), plotted as orbs representing the 12 ROIs. A) Orbs plotted at the centroid of each of our 12 ROIs and size scaled across frequencies based on proportion of time within each task state. B) Orb size now scaled by regional contribution to proportion of time in each task state for each frequency, as indicated by the colored boxes and labels.

## Discussion

We investigated the directionality in brain-behavior interactions during a simulated driving task, and identified two distinct neuro-behavioral states. The Proactive state consists of task intervals when brain activity precedes behavior and seen as the cause of the behavior. The Proactive state is flexible and dependent on task demands, is most predictive within the delta and beta bands, and may reflect motor execution and error determination. The Reactive brain state, on the other hand, appears inflexible to task demands, is most predictive by posterior parietal-motor-frontal alpha band activity, and may reflect the monitoring of environmental statistics. More generally, our results indicate the power of this neuro-behavioral method that is not constrained by segregation and averaging over experimental trials; instead, this neuro-behavioral analysis reveals the relationship between neural signals and continuous behavioral measurements, enabling the study of dynamic fluctuations in task performance dependent on idiosyncratic changes in neuro-behavioral state.

*The Proactive Neuro-behavioral State*

The Proactive state, where brain activity precedes and predicts behavior, is characterized by diffuse anterior-posterior delta and beta oscillatory activity, and it seems to reflect a state which *actively* plans the response to the sensory information. Subjects were rarely in this state within the experiment, most likely due to the fact that little response was necessary to maintain course. However, this state was more predominant for steering behavior than heading error, and this likely reflects demand characteristics of the driving measurement since steering is arguably a more *proactive* behavioral estimate than heading error. This result was captured in the histogram analysis with a larger number of windows for significant timeframes in the Proactive state as well as the analysis of transition probabilities. We observed increased transitions from Proactive-to-Reactive and Reactive-to-Proactive states in conjunction with steering behavior. These *between* state transitions may suggest that subjects quickly switch to the Proactive brain state when an action is needed. The flexibility of this state is also supported by the interaction between state and our two naturalistic driving conditions that revealed more flexibility in the Proactive state dependent on environmental statistics.

Regions and frequency bands also support the notion that the Proactive state is, indeed, one of action. Numerous studies have associated beta band activity generated in motor cortex and surrounding areas with preparatory action (Alegre et al., 2003; Baker, 2007; Tan et al., 2013; Tzagarakis et al., 2015). Beta activity of this sort is also coherent with electromyographic activity (Baker et al., 1999) and is significantly linked with BOLD fluctuations within motor and pre-motor regions (Ritter et al., 2009). Although delta band activity is often associated with sleep (Amzica and Steriade, 1998), it has also been prominent in the decision making literature, including making judgments to discriminate stimuli in auditory oddball tasks (Başar-Eroglu et al., 1992; Schürmann et al., 1995). Moving beyond these behavioral links, the topographic distribution of delta band activity is often diffuse across scalp electrodes, and previous research has interpreted this diffuse pattern as consistent with a "distributed response system" (Başar et al., 2001). Our source analysis also finds a diffuse anterior-posterior network that complements this previous literature. Collectively, our results support the notion that the Proactive brain state is one of preparation and action, including a role in deciding and planning the response.

The posterior cingulate cortex was a strong contributor of the Proactive state within the delta band, but it may play a more regulating role within this network. It is the primary hub in the default mode network, a network of brain regions that show reliable deactivation during a variety of cognitive tasks (Raichle et al., 2001), and recent research has shown it to play a more active role in regulating cognition (Gilbert and Sigman, 2007; Hampson et al., 2006; Leech et al., 2012; Pearson et al., 2011). In the nonhuman primate, the PCC has also been shown to signal environmental change and the need to alter behavior (Hayden et al., 2010), so this role would also be relevant within the driving task.

These results raise an interesting question about whether these two brain networks, separable by frequency bands and regional contribution, communicate with one another to accomplish task aims. Though beyond the scope of the current study, there is substantial evidence that relationships between fast and slow rhythms in the brain are linked to behavioral action (or inaction) under varying levels of motivation (Putman, 2011; Schutter and Van Honk, 2005).

*The Reactive Neuro-behavioral State*

Participants spent most of the 45 minute, simulated driving session in a Reactive state, where driving performance behavior predicts brain activity. The Reactive state seems to process incoming information and *react* to environmental statistics. We attribute the predominance of this state to high monotony in the task. Consistent with a reactive interpretation, this neuro-behavioral state was predominantly associated with alpha activity in posterior parietal, motor, and frontal regions. Alpha band activity (8-12Hz) is an intrinsic brain oscillation of fervent study due to its prominence in resting EEG and sensitivity to various task demands. Several hypotheses have been proposed ascribing a functional role to its presence in EEG. The first was proposed by Adrian and Mathews (1934) who found that the power within the alpha band increases when subjects are awake with eyes closed. They interpreted this as alpha band activity reflecting a brain state of inactivity, priming the brain for incoming information. This theory has been expanded and revised to more clearly represent 'cortical idling' (Pfurtscheller et al., 1996), and modern extensions of this have shown that even at a shorter temporal scale, alpha activity may gate perceptual information (Jensen and Mazaheri, 2010; van Dijk et al., 2008). More recently, however, these theories have been further developed, proposing that alpha activity represents controlled access to a knowledge system, constrained by the limits of attention (Klimesch et al., 2007). Within the context of this driving task, a reactive state where behavior predicts brain activity appears to be consistent with these theories about alpha oscillations, and we interpret its role as gating perceptual information, allowing the mind to wander until a salient perceptual event that requires a behavior response occurs in the environment and requires the brain to react and correct performance.

Results from both the asymmetry values and transition probabilities demonstrated that this Reactive state was sustained, and our analysis of the two naturalistic driving conditions revealed inflexibility in the Reactive brain state since it is not dependent on environmental statistics. The underlying network consisted mostly of sensory regions and attentional regions in parietal cortex. In our experiment, these regions contributed equally to this neuro-behavioral state and suggests an interesting network of attentional sensory gating. In particular, the occipital and parietal regions are regularly implicated in visual perception and attention tasks, and they have been consistently shown to be sources of alpha band activity(e.g., Laufs et al., 2003).

Together, these findings may characterize a Reactive neuro-behavioral state associated with an alpha-band, sensory gating network with attentional constraints to indicate environmental change. Future research applying this method to more dynamic environments and a diverse set of tasks will determine the flexibility of this sensory-driven brain state. It would also be interesting to examine the connectivity between the regions supporting this neuro-behavioral state.

*Advancements in linking behavior with brain activity*

Although this analysis focused on behavior-brain dynamics while driving, this approach can be used more generally to study relationships between functional brain networks and continuous task performance in other domains. Examining the directional relationship between brain and behavior operationalizes the concept of a brain state, emphasizing the study of large-scale oscillatory activity from EEG data to investigate cross-region communication and whole-brain dynamics. Thus, it can test the hypothesis that synchronized frequency oscillations provide a mechanism to form transient networks that can integrate information across local, specialized brain regions (He et al., 2015; Klimesch et al., 2007), and it can reveal how specific oscillations result in behavioral consequences and dynamic fluctuations in task performance. Here, we identified relationships between the temporal dynamics of EEG oscillatory activity and continuous fluctuations in two behavioral measures of driving performance in a simulated environment, the participant's steering behavior and vehicle's heading error. The directional influence reveals insights about the role of neuro-behavioral states, indicating when an individual is actively planning a course of action versus timeframes when the person is merely reacting to salient events in the environment. Future research will determine whether brain-behavior interactions found within additional task domains also show controlled transitions among brain states in conjunction with ongoing task demands.

**Acknowledgements**

We would like to thank Piotr Franaszczuk and Nima Bigdely Shamlo for input on analyses and earlier drafts of this work and Patrick Connolly for data collection. Research was sponsored by the U.S. Army Research Laboratory, including work under Cooperative Agreement Numbers W911NF-10-2-0022. The views and conclusions contained in this document are those of the authors and should not be interpreted as representing the official policies, either expressed or implied, of the Army Research Laboratory or the U.S. Government.